\documentclass[10pt,letterpaper]{IEEEtran}
\usepackage[latin9]{inputenc}
\usepackage{units}
\usepackage{amsmath}
\usepackage{amssymb}
\usepackage{graphicx}

\makeatletter

\usepackage{cite}

\usepackage[printonlyused]{acronym}
\acrodef{QPSK}{quadrature phase-shift keying}
\acrodef{NFDM}{nonlinear frequency-division multiplexing}
\acrodef{NFT}{nonlinear Fourier transform}
\acrodef{FNFT}{forward NFT}
\acrodef{BNFT}{backward NFT}
\acrodef{OFDM}{orthogonal frequency-division multiplexing}
\acrodef{TX}{transmitter}
\acrodef{RX}{receiver}
\acrodef{FT}{Fourier transform}
\acrodef{DAC}{digital-to-analog converter}
\acrodef{ADC}{analog-to-digital converter}
\acrodef{LP}{Layer-Peeling}
\acrodef{SNR}{signal-to-noise ratio}
\acrodef{NIS}{nonlinear inverse synthesis}
\acrodef{DBP}{digital backpropagation}
\acrodef{QAM}{quadrature amplitude modulation}
\acrodef{SE}{spectral efficiency}
\acrodef{EDC}{electronic dispersion compensation}
\acrodef{NLSE}{nonlinear Schrödinger equation}
\acrodef{AWGN}{additive white gaussian noise}

\makeatother

\begin{document}

\title{Why Noise and Dispersion may Seriously Hamper Nonlinear Frequency-Division
Multiplexing\\
}

\author{Stella~Civelli,~Enrico~Forestieri, \IEEEmembership{Member, IEEE},~and~Marco~Secondini,
\IEEEmembership{Member, IEEE}\thanks{S. Civelli, E. Forestieri, and M. Secondini are with the Tecip Institute,
Scuola Superiore Sant'Anna, Pisa, Italy and with the National Laboratory
of Photonic Networks, CNIT, Pisa, Italy (e-mail: stella.civelli@santannapisa.it).}\thanks{Part of this work was presented at the Munich Workshop on Information
Theory of Optical Fiber (MIO) 2016 and at the Progress In Electromagnetics
Research Symposium (PIERS) 2017, Saint Petersburg, Russia. }\thanks{This work was supported in part by Fondazione Cassa di Risparmio di
Firenze, under the grant NOSTRUM: NOnlinear SpecTRUm Modulation.}}
\maketitle
\begin{abstract}
The performance of optical fiber systems based on \ac{NFDM} or on
more conventional transmission techniques is compared through numerical
simulations. Some critical issues affecting \ac{NFDM} systems\textemdash namely,
the strict requirements needed to avoid burst interaction due to signal
dispersion and the unfavorable dependence of performance on burst
length\textemdash are investigated, highlighting their potentially
disruptive effect in terms of spectral efficiency. Two digital processing
techniques are finally proposed to halve the guard time between NFDM
symbol bursts and reduce the size of the processing window at the
receiver, increasing spectral efficiency and reducing computational
complexity.
\end{abstract}


\section{Introduction}

\IEEEPARstart{N}{onlinear} frequency-division multiplexing (NFDM)
has recently attracted attention as a way to cope with nonlinear effects
in optical fiber communications \cite{Yousefi2014_NFT,le2014nonlinear,le2016,turitsyn2017optica}.
By using the \ac{NFT} the optical signal can be represented through
its nonlinear spectrum, whose evolution along the fiber is governed
by a simple linear equation \cite{Yousefi2014_NFT}. This is exploited
in \ac{NFDM} systems by encoding information directly onto the nonlinear
spectrum, such that it can be easily retrieved at the receiver avoiding
nonlinear interference due to propagation.

Different flavors of \ac{NFDM} do exist, depending on the considered
\ac{NFT} boundary conditions and on the way information is mapped
onto the nonlinear spectrum \cite{turitsyn2017optica}. So far, vanishing
boundary conditions have been mostly used \cite{le2014nonlinear,Yousefi2014_NFT},
the only exception (to the best of our knowledge) being the periodic
boundary conditions employed in \cite{kamalian2016periodic}. With
vanishing boundary conditions, the nonlinear spectrum has a continuous
part, analogous to the linear spectrum, and some discrete components
(solitons) arising for specific input profiles and with no linear
counterpart. Accordingly, different \ac{NFDM} schemes have been proposed
in which information is encoded on the continuous spectrum \cite{le2014nonlinear},
discrete spectrum \cite{Yousefi2014_NFT}, or both~\cite{Aref2016}.

Despite many recent theoretical and experimental publications on
the subject, a clear indication of whether \ac{NFDM} can actually
outperform conventional systems is still missing in the literature.
Moreover, we believe that the impact of some potentially critical
issues has been overlooked. Thus, we compare here the performance
of \ac{NFDM} systems with that of conventional ones employing ideal
\ac{EDC} or \ac{DBP}, focusing on the strict limitations imposed
by temporal broadening (due to fiber dispersion) and nonlinear spectrum
perturbation (due to amplifier noise) to NFDM systems. We consider
vanishing boundary conditions and modulation of the continuous spectrum,
following the \ac{NIS} approach \cite{le2014nonlinear,le2016}. The
first choice is due to the higher simplicity and maturity of the underlying
theory; the second one is due to our belief that the modulation of
the continuous spectrum is essential to achieve high spectral efficiencies.\footnote{The \ac{NIS} technique is a nonlinear analogous of \ac{OFDM}, tending
to it in the linear regime \cite{le2014nonlinear}. Thus, it can be
combined with conventional coding and modulation to achieve high spectral
efficiencies and approach channel capacity in the linear regime.} 

Nevertheless, as we will show in the following, also the \ac{NIS}
technique suffers from some important drawbacks that, if not clearly
understood and overcome, may severely limit performance and spectral
efficiency. In fact, transmission is organized in bursts of $N_{b}$
information symbols, separated by a guard time ($N_{z}$ spaces) to
avoid burst interaction during propagation. The guard time: \emph{i)}
plays a role similar to the cyclic prefix in \ac{OFDM} (emulating,
at least within the limit of channel memory, the boundary conditions
required by the underlying theory); \emph{ii)} should at least equal
the maximum time broadening induced by fiber dispersion; and \emph{iii)}
causes a reduction of the overall spectral efficiency by a factor
$\eta=N_{b}/(N_{z}+N_{b})$\textemdash in the following referred to
as \emph{rate} \emph{efficiency}. We show by numerical simulations
that the simple solution (used in conventional \ac{OFDM}) of increasing
$N_{b}$ to limit the loss of spectral efficiency is not feasible,
because the performance decreases with $N_{b}$ due to a sort of signal-noise
interaction taking place at the receiver. This behavior is in agreement
with a recent analytical model \cite{Turitsyn_nature16}, and is substantially
different from the behavior of conventional systems, whose performance
saturates to a finite value when $N_{b}\rightarrow\infty$ ($\eta\rightarrow1$).
Moreover, the computational complexity of the \ac{NFT} remains a
major concern for the implementation of NFDM systems, despite some
recent progresses toward its reduction (see \cite{turitsyn2017optica}
and references therein). With this in mind, we investigate two techniques
aimed at mitigating the impact of time broadening on spectral efficiency
and computational complexity. We show, through numerical simulations,
that a precompensation technique can halve the necessary guard time
between different bursts, and that a windowing technique can significantly
shorten the length of the processed signal without performance degradation
or even provide a small gain.

\begin{figure}
\begin{centering}
\includegraphics[width=1\columnwidth]{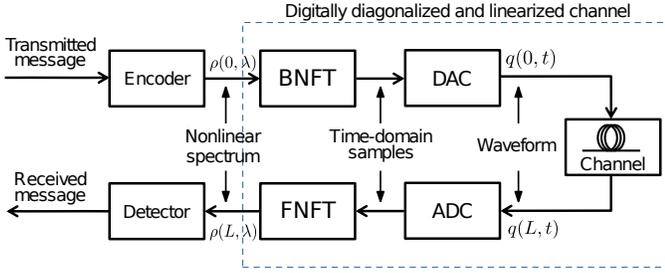}\vspace*{-1ex}
\par\end{centering}
\caption{\label{fig:1}NFDM transmission scheme}
\end{figure}

\section{System performance }

The considered \ac{NFDM} transmission scheme is sketched in Fig.\,\ref{fig:1}
and is based on the \ac{NIS} technique proposed in \cite{le2014nonlinear}.
The \ac{TX} encodes the information on a \ac{QPSK} signal, whose
\ac{FT} is mapped onto the continuous part of the input nonlinear
spectrum $\rho(0,\lambda)$. The \ac{BNFT} block generates, through
a \ac{DAC}, the corresponding optical signal $q(0,t)$. While $q(z,t)$
evolves along the fiber according to the \ac{NLSE}, its nonlinear
spectrum evolves linearly according to $\rho(z,\lambda)=\rho(0,\lambda)e^{-j4\lambda^{2}z}$
\cite{Yousefi2014_NFT}. At the \ac{RX}, the output optical signal
$q(L,t)$ is sampled by the \ac{ADC} and sent to the \ac{FNFT} block,
which computes the corresponding output nonlinear spectrum $\rho(L,\lambda)$
(corrupted by amplifier noise during propagation). Finally, the detector
multiplies $\rho(L,\lambda)$ by $e^{j4\lambda^{2}L}$ to remove the
propagation effect; performs an inverse \ac{FT}, followed by matched
filtering and symbol-time sampling to obtain a noisy replica of the
transmitted symbols; and makes decisions based on a minimum Euclidean
distance criterion.\footnote{Besides its simplicity and widespread use, this detection strategy
achieves the capacity bound in \cite{Turitsyn_nature16} and is asymptotically
optimal at low power. However, a better performance might be achievable
by a more accurate detection strategy, accounting for the actual statistics
of the received nonlinear spectrum.}

The \ac{QPSK} signal power spectral density is raised-cosine shaped
with roll-off factor $\beta=0.2$ (a typical choice in conventional
systems), while the symbol rate is $R_{s}=1/T_{s}=\unit[50]{GBd}$.
The physical channel is a standard single-mode fiber of length $L=\unit[2000]{km}$,
attenuation $\alpha=\unit[0.2]{dB/km}$, dispersion $\beta_{2}=\unit[-20.39]{ps^{2}/km}$,
and nonlinear coefficient $\gamma=\unit[1.22]{W^{-1}km^{-1}}$, along
which ideal distributed amplification with spontaneous emission factor
$\eta_{\mathrm{sp}}=4$ is considered. The bandwidth of both the \ac{DAC}
and the \ac{ADC} is $\unit[100]{GHz}$. The \ac{BNFT} is computed
by an enhanced version of the Nyström method \cite{civelliNFT}, while
the \ac{LP} method is employed for the \ac{FNFT} \cite{Yousefi2014_NFT,le2014nonlinear}.
An oversampling factor of $4$ samples per symbol was considered for
both \ac{BNFT} and \ac{FNFT}, unless otherwise specified. At high
powers, some rare but disruptive numerical instabilities in the calculation
of the noisy nonlinear spectrum were observed \cite{Civelli_fotonica16}.
We conjecture they are related to the rise of discrete components
in the nonlinear spectrum (solitons) induced by noise \cite{bulow2015experimental}.
This issue has been practically resolved by resorting to linear interpolation
between adjacent frequencies when the phenomenon occurs at a given
frequency. As explained later, this numerical expedient is not required
when employing the windowing technique proposed in the next section.

\begin{figure}
\quad{}\raisebox{20ex}{(a)}\quad{}\includegraphics[width=0.8\columnwidth]{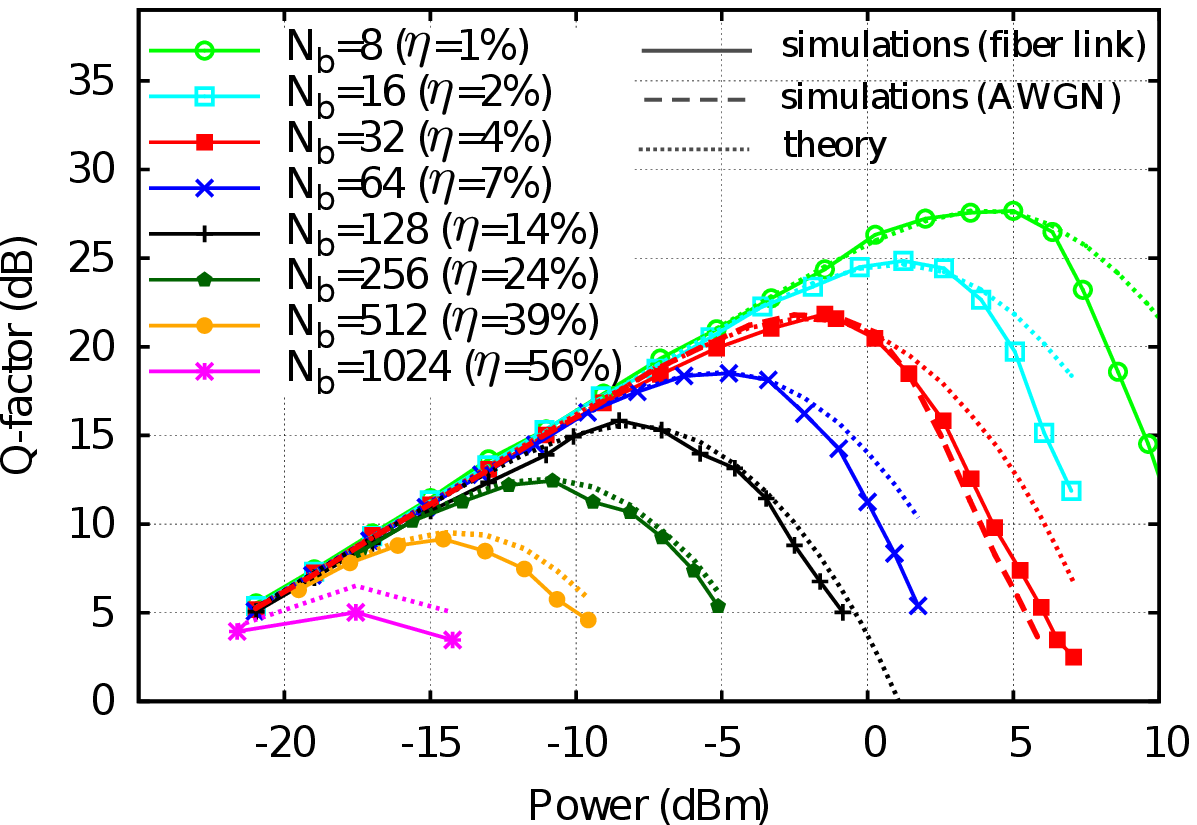}\vspace*{-2ex}
\qquad{}

\quad{}\raisebox{20ex}{(b)}\quad{}\includegraphics[width=0.8\columnwidth]{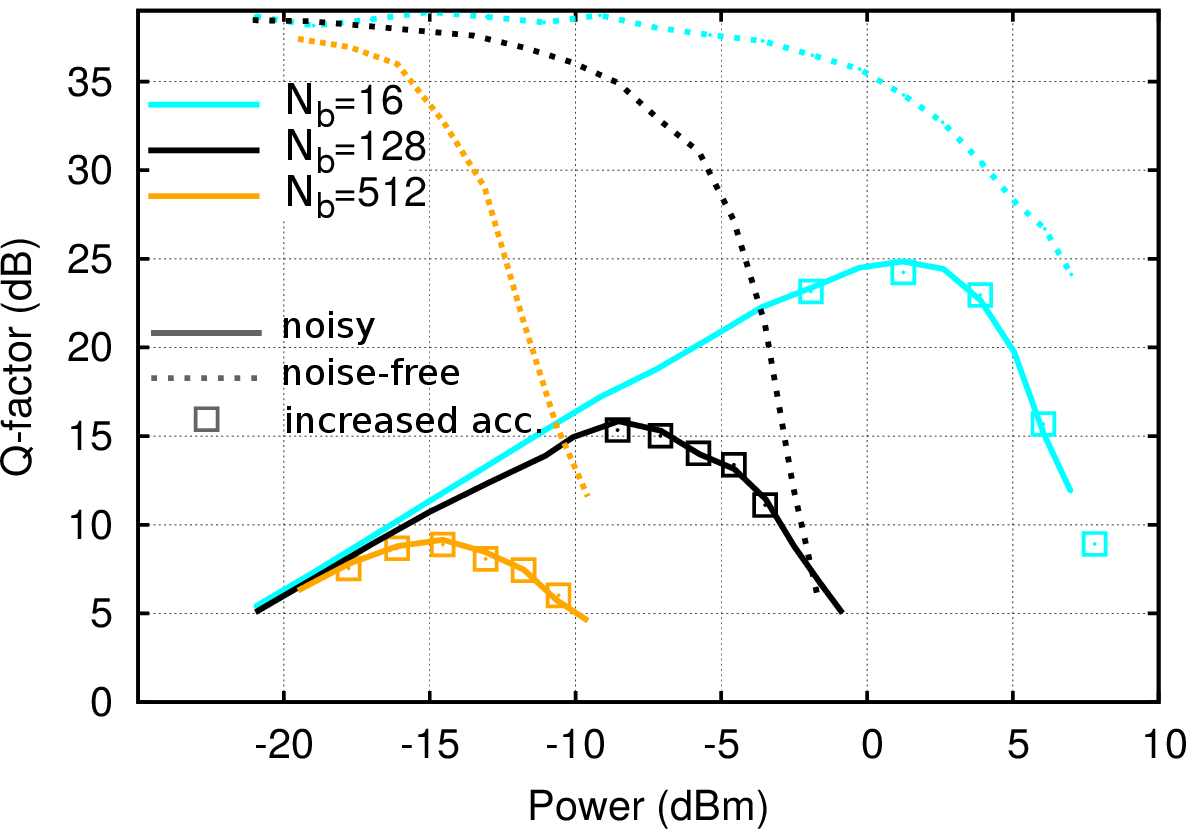}\vspace*{-2ex}

\caption{\label{fig:2uno}(a) Q-factor vs optical power for standard \protect\ac{NFDM}
with different burst length $N_{b}$ (and rate efficiency $\eta$);
(b) Impact of numerical inaccuracies on the Q-factor.}
\vspace*{-2ex}
\end{figure}

\begin{figure}
\quad{}\raisebox{20ex}{(a)}\quad{}\includegraphics[width=0.8\columnwidth]{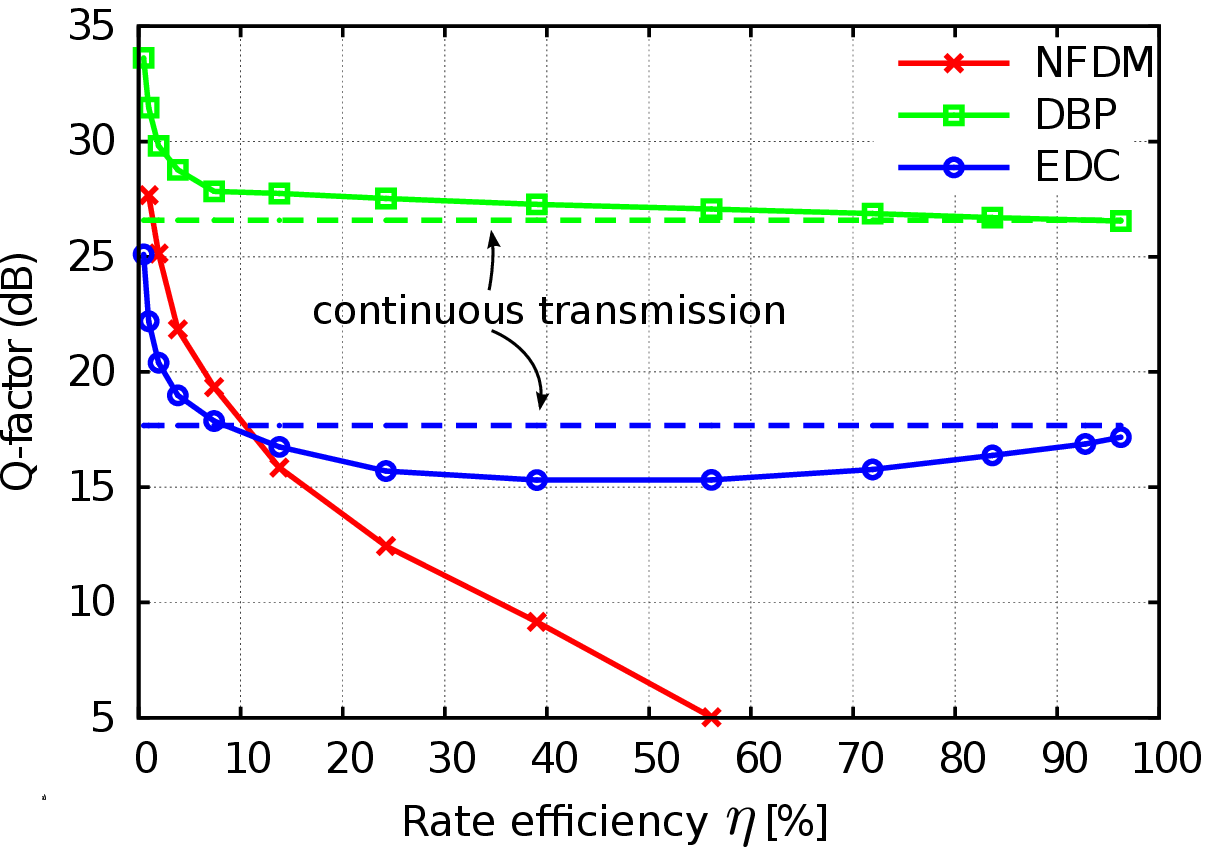}\vspace*{-2ex}
\qquad{}

\quad{}\raisebox{20ex}{(b)}\quad{}\includegraphics[width=0.8\columnwidth]{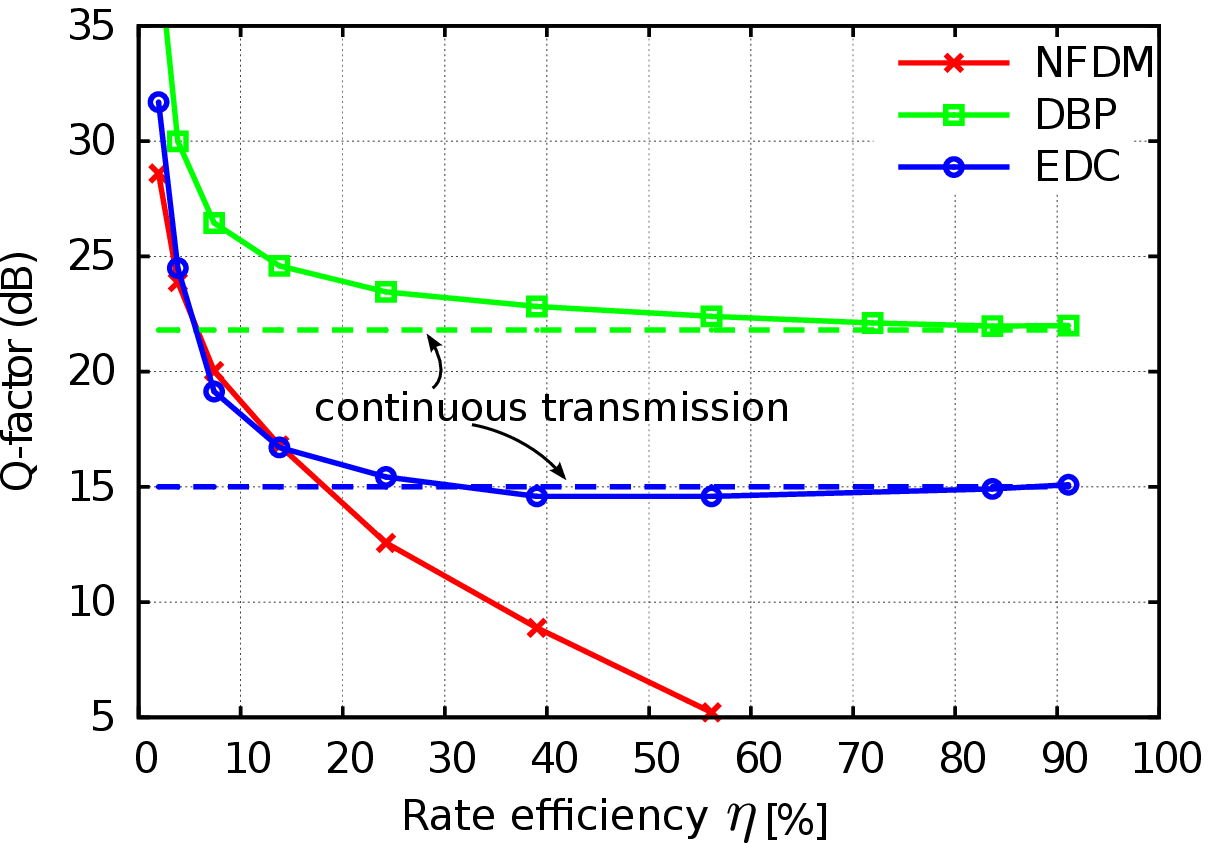}\vspace*{-2ex}

\caption{\label{fig:2due}(a) optimal Q-factor vs rate efficiency for \protect\ac{NFDM}
and conventional systems; (b) optimal Q-factor vs rate efficiency
for \protect\ac{NFDM} and conventional systems with $\beta_{2}=\unit[-1.27]{ps^{2}/km}$. }
\vspace*{-2ex}
\end{figure}

Fig.~\ref{fig:2uno}(a) shows the Q-factor from numerical simulations
(solid lines) or from the effective \ac{SNR} of the analytical model
in \cite{Turitsyn_nature16} (dotted lines), for different burst lengths
$N_{b}$ and a guard interval of $N_{z}=800$ zeros. As required,
this value is slightly larger than the overall channel memory set
by the time broadening induced by dispersion, which is of the order
of $2\pi|\beta_{2}|LR_{s}^{2}(1+\beta)$ symbol times ($\sim$768
in our case)\footnote{The initial broadening induced by the BNFT at the receiver is usually
negligible compared to that induced by dispersion, as shown in \cite{Civelli_fotonica16}.
This is not true for the lowest curve of Fig.~\ref{fig:2uno}(a)
($N_{b}=1024$), such that an additional penalty is observed compared
to the analytical model.}. The corresponding rate efficiencies $\eta$ are also indicated.
After reaching a maximum at some optimum power, all curves fall down
since the impact of amplifier noise on the nonlinear spectrum increases
with signal energy (a sort of signal-noise interaction). The impact
of numerical inaccuracies, already considered in \cite{Civelli_fotonica16},
is shown in Fig.\,\ref{fig:2uno}(b), where the simulation results
of Fig.\,\ref{fig:2uno}(a)  are compared with the results obtained
in the corresponding noise-free scenario  and with those obtained
with higher sampling rate ($16$ samples per symbol) and longer guard
time ($N_{z}=900$ guard symbols). It is apparent that, in the region
near the optimal power, the noise-free curves are above the noisy
ones; moreover, the performance remains unchanged if a higher accuracy
is considered. Therefore, Fig.\,\ref{fig:2uno}(b) confirms that
the observed performance degradation is due to the interaction of
signal and noise, rather than to numerical inaccuracies. The agreement
between theory and simulations up to the optimum power further confirms
that the obtained results are not affected by limitations of the numerical
algorithms. Moreover, it verifies the accuracy of the perturbation
approach and asymptotic approximations used in \cite{Turitsyn_nature16}
for the computation of the effective SNR. Remarkably, the maximum
reduces as $N_{b}$ increases. This behavior persists for bursts longer
than the channel memory, as shown for $N_{b}=1024$. Interestingly,
unlike conventional systems, the performance of the \ac{NFDM} scheme
considered here remains unchanged (at least in the considered range
of powers) if the optical fiber channel is replaced with an \ac{AWGN}
channel with same accumulated noise (shown with dashed line for $N_{b}=32$;
results are similar for any $N_{b}$). We will return on this later.

To better understand this issue and its relevance, the maximum of
each curve in Fig.\,\ref{fig:2uno}(a) is reported in Fig.\,\ref{fig:2due}(a)
as a function of the rate efficiency $\eta$, along with the corresponding
performance of ideal \ac{DBP} and \ac{EDC} methods. For the sake
of comparison, burst mode transmission with $N_{z}=800$ and same
modulation parameters were considered in all cases. As expected, the
performance of both \ac{DBP} and \ac{EDC} converges to that of a
continuous transmission for bursts longer than the channel memory
($N_{b}>N_{z}$, corresponding to $\eta>0.5)$. This is because, in
these systems, nonlinear interaction involves only signal and noise
components that are closer in time than the overall channel memory.
On the other hand, the \ac{NFDM} performance keeps decreasing even
for longer bursts, as in this case signal-noise interaction does occur
during fiber propagation, as in conventional systems, but at the receiver
over the full integration window used for computing the nonlinear
spectrum (the \ac{FNFT}), with an impact that increases with the
total signal energy therein. This is confirmed by the curve shown
in Fig.\,\ref{fig:2uno}(a) for the \ac{NFDM} system over an \ac{AWGN}
channel. The comparison in Fig.\,\ref{fig:2due}(a) also reveals
that ideal \ac{DBP} performs better than \ac{NFDM}, which performs
better than \ac{EDC} only for bursts of short-medium length. 

In the single-user scenario considered here, its worse performance
with respect to ideal \ac{DBP} may be not a crucial issue, as \ac{NFDM}
is expected to perform better in multi-user scenarios, where \ac{DBP}
is much less effective because of inter-channel interference \cite{Yousefi2016}.
A more critical issue is the peculiar dependence of its performance
on burst length, as also confirmed by theory. In fact, as shown in
Fig.\,\ref{fig:2due}(a), a reasonable performance is obtained only
at the expense of a low rate efficiency. For $\eta>0.11$, \ac{NFDM}
performs worse than simple \ac{EDC}, definitely loosing any appeal.
Moreover, also the computational complexity of most practical \ac{NFT}
algorithms has an unfavorable dependence on the total signal length
(burst length plus guard time). Note that, if there were no broadening,
a much shorter guard time and burst length could be considered, with
a significant improvement in performance and complexity. Thus, one
is tempted to check whether it is any better in links with low-dispersion
fibers. Considering a sixteen times lower dispersion ($\beta_{2}=\unit[-1.27]{ps^{2}/km}$)
and guard time ($N_{z}=50)$, Fig.\,\ref{fig:2due}(b) shows that
the dependence of \ac{NFDM} performance on the rate efficiency remains
practically unchanged (signal-noise interaction in the \ac{FNFT}
is reduced for a shorter burst, but is increased for a lower dispersion,
the two effects canceling out). Instead, \ac{DBP} and \ac{EDC} performance
worsens at high rates and slightly improves at low rates. The overall
picture does not change significantly: \ac{DBP} outperforms \ac{NFDM}
and \ac{EDC}, whose performance is almost the same up to $\eta=0.14$.
For $\eta>0.14$, \ac{NFDM} performance degrades much faster than
\ac{EDC} and \ac{DBP}, as in the previous scenario.

\begin{figure}
\centering{}\includegraphics[width=0.8\columnwidth]{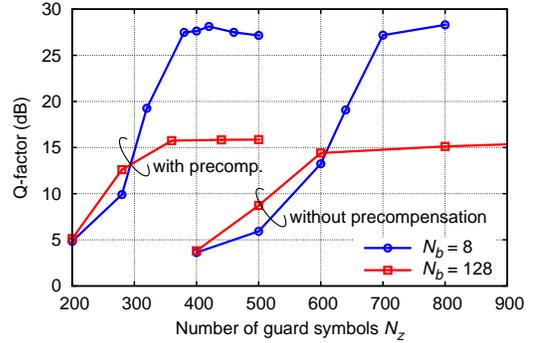}\vspace*{-2ex}
\caption{\label{fig:3} Optimal Q-factor vs guard interval with and without
precompensation. }
\vspace*{-2ex}
\end{figure}

\section{Precompensation and windowing}

The large number of guard spaces $N_{z}$ required to avoid burst
interaction during propagation reduces the transmission rate by the
factor $\eta=N_{b}/(N_{z}+N_{b})$, with a significant loss of spectral
efficiency. For mitigating this loss, we introduce a precompensation
technique that, by minimizing the time broadening induced by dispersion
throughout the link, allows to reduce $N_{z}$.\footnote{The same technique has been independently proposed in \cite{tavakkolnia2016}.}
In order to remove the propagation effect from the received nonlinear
spectrum $\rho(L,\lambda)$, rather than multiplying it by $e^{j4\lambda^{2}L}$
as done in standard \ac{NIS}, we split the compensation between \ac{TX}
and \ac{RX}, both pre-multiplying $\rho(0,\lambda)$ and post-multiplying
$\rho(L,\lambda)$ by $e^{j2\lambda^{2}L}$. This is the same as generating
the signal at a distance $-L/2$ and propagating it to a distance
$L/2$. In this way, the same time broadening of $\pi|\beta_{2}|LR_{s}^{2}(1+\beta)$
symbol times is observed at \ac{TX} and \ac{RX}, in fact halved
with respect to the standard implementation.

Fig.~\ref{fig:3} shows the Q-factor obtained with and without precompensation
for bursts of length $N_{b}=8$ and $N_{b}=128$ at their optimal
launch power (about $\unit[3.8]{dBm}$ and $\unit[-8.5]{dBm}$, respectively)
as a function of the number of guard symbols $N_{z}$. For both burst
lengths, precompensation allows using half the guard time to achieve
the same performance, with a significant increase of the rate efficiency
(almost doubled). This precompensation technique reduces the computational
complexity of the \ac{FNFT} while increasing the \ac{BNFT} one (as
the total processing windows are, respectively, shortened and lengthened),
with an overall effect that depends on the algorithms employed for
the \ac{BNFT} and \ac{FNFT}.

\begin{figure}
\quad{}\raisebox{20ex}{(a)}\negmedspace{}\includegraphics[width=0.9\columnwidth]{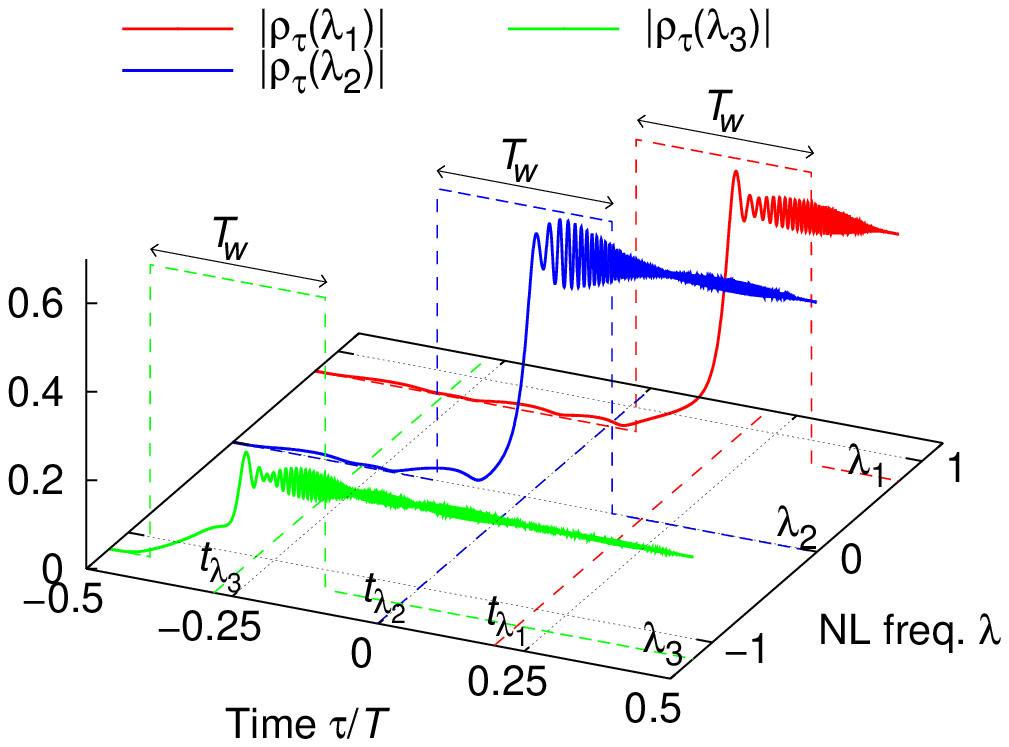}\vspace*{-2ex}
\qquad{}

\quad{}\raisebox{20ex}{(b)}\qquad{}\includegraphics[width=0.7\columnwidth]{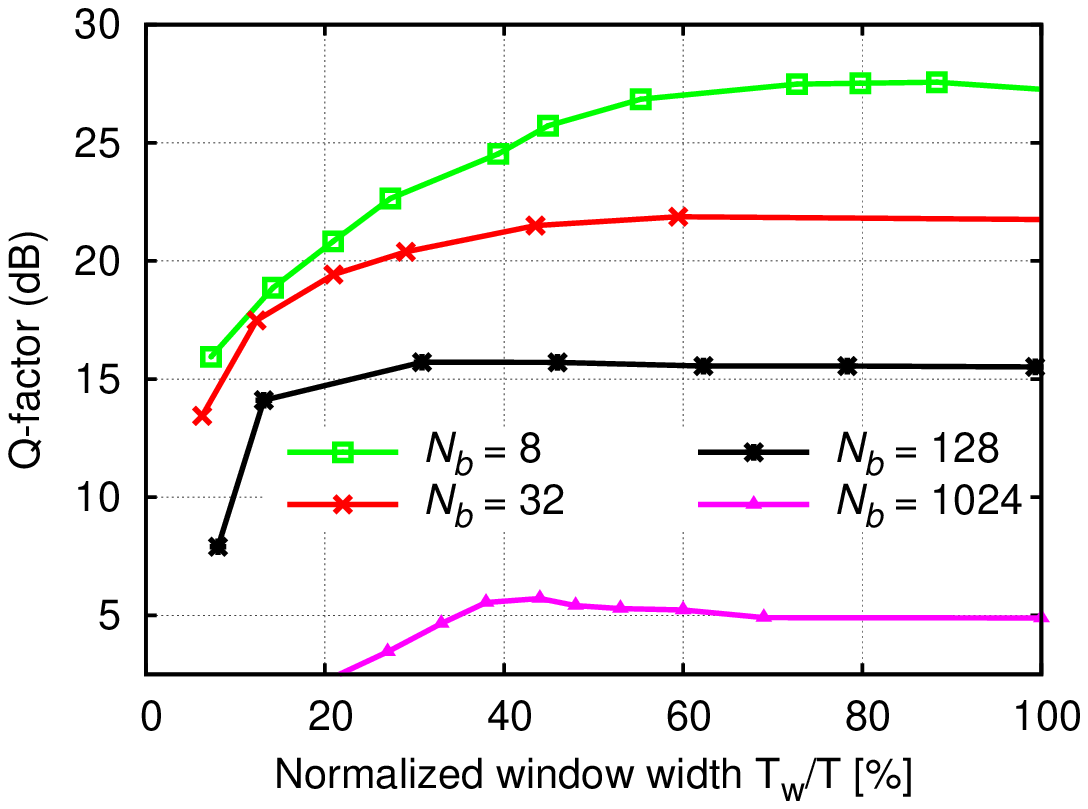}\vspace*{-2ex}

\caption{\label{fig:4}(a) Modulus of the noiseless nonlinear spectrum (vertical
axis) vs upper time limit considered in the \protect\ac{LP} algorithm
($\tau$ axis) for different spectral components; (b) optimal Q-factor
when using the windowing technique vs window width.}
\vspace*{-2ex}
\end{figure}

As mentioned above, time broadening affects also the computational
complexity of the \ac{FNFT} at \ac{RX}, which, in principle, must
be performed for each burst on the entire time range $-T/2<t<T/2$,
with $T=(N_{b}+N_{z})T_{s}$. Nevertheless, some computational savings
can be achieved by noting that, similarly to the linear spectrum,
also the continuous part of the nonlinear spectrum experiences a sort
of group velocity dispersion during propagation, with different frequency
components traveling at different speeds. As a result, different time
portions of the received optical signal bring information about different
spectral components of the nonlinear spectrum. This is illustrated
in Fig.\,\ref{fig:4}(a), which shows the modulus of the nonlinear
spectrum (vertical axis) as obtained when applying the \ac{LP} algorithm
to the received optical signal truncated to the time interval $-T/2<t<\tau$,
with the upper limit reported on the $\tau$ axis. Three different
spectral components $\lambda$ are reported at different depths in
the graph. For illustration purposes, results are shown in the absence
of optical noise. It is apparent that, for each spectral component
$\lambda$, only a small portion of the received optical signal\textemdash contained
in a time window whose center depends linearly on the considered frequency\textemdash contributes
to the final value of the nonlinear spectrum. This suggests the following
windowing technique: given the received optical signal, each frequency
component $\rho(L,\lambda)$ is computed by applying the \ac{LP}
method on the moving time window $\mathrm{max}\{t_{\lambda}-T_{w}/2,-T/2\}<t<\mathrm{min}\{t_{\lambda}+T_{w}/2,T/2\}$,
where $T_{w}<T$ is the window width (to be optimized) and $t_{\lambda}=-2\beta_{2}L\lambda/T_{0}$
its center, $T_{0}$ being the time normalization parameter used to
define the NFT \cite{Yousefi2014_NFT}.

Fig.~\ref{fig:4}(b) shows the Q-factor at optimum power obtained
by the described windowing technique as a function of the window width
$T_{w}$, for the same system in Fig.~\ref{fig:2uno}(a). These
results show that the time window for computing the \ac{FNFT} can
be reduced to about 70\%, 50\%, 20\%, and 40\% of the total signal
duration for $N_{b}=8$, $32$, $128$, and $1024$, respectively,
with significant computational savings. The different behavior for
different burst lengths depends on the maximum achievable Q-factor
and on the initial burst length: the lower the Q-factor, the narrower
the time window where the signal contribution dominates over noise,
until the window width becomes much smaller than $T_{s}N_{b}$. 

Moreover, this technique avoids the excess noise outside the window
of interest for each considered frequency, slightly improving performance
(as shown for $N_{b}=1024$ in Fig.~\ref{fig:4}(b)) and reducing
the numerical instabilities of the \ac{LP} algorithm mentioned in
the previous section. In fact, when using the windowing technique,
we were able to reproduce the same results of Fig.~\ref{fig:2uno}(a)
without the need to resort to the interpolation expedient and almost
avoiding the small penalty (compared to the theoretical curves) observed
in Fig.~\ref{fig:2uno}(a) for $N_{b}=1024$ near the optimum launch
power.

\section{Conclusions}

Some critical issues arising in \ac{NFDM} systems based on vanishing
boundary conditions and modulation of the continuous spectrum have
been investigated. According to current theory, these systems operate
in burst mode, with a guard time between bursts long enough to accommodate
the temporal broadening induced by the accumulated dispersion. Our
findings show that, unlike in conventional systems, in the NIS scheme
considered in this work, the spectral efficiency loss due to the guard
time cannot be reduced by increasing at will the burst length. In
fact, the performance degrades as the burst length increases due to
a detrimental signal-noise interaction taking place in the \ac{FNFT}
operation at the receiver. The impact of this effect is such that,
in a 50~GBd \ac{QPSK} \ac{NIS} system over 2000~km of standard
fiber, the effective rate cannot be increased to more than 11~Gb/s
(about 100 information symbols and 800 guard symbols) without degrading
its performance below that of a conventional system with simple \ac{EDC}.
Similar results are found when considering low-dispersion fibers ($D=1$~ps/nm/km).
A digital precompensation technique has been proposed to halve the
guard time required between different bursts of information symbols,
allowing to almost double the \ac{NFDM} spectral efficiency. Moreover,
a windowing technique has been introduced to limit the computational
complexity at the \ac{RX} and to avoid excess noise entering the
\ac{FNFT} computation in the presence of temporal broadening. These
improvements, though significant, are still not sufficient to make
\ac{NIS} an attractive replacement to conventional systems. In order
to pave the way for the advent of NFDM systems and exploit their great
potentials \cite{Yousefi2016}, the critical issue highlighted in
this work should be solved, for instance by devising more appropriate
detection strategies based on the actual statistics of the noisy nonlinear
spectrum.

\bibliographystyle{ieeetr}
\bibliography{ref}

\end{document}